\documentclass[useAMS,usenatbib]{mn2e}

\usepackage[dvips]{graphicx}
\usepackage{amsmath}
\usepackage{amssymb}
\usepackage{gensymb}

\title{Discovery of Carbon Radio Recombination Lines in absorption towards Cygnus~A}
\author[J.~B.~R. Oonk et al.]{J.~B.~R.~Oonk$^{1}$\thanks{E-mail:
oonk@astron.nl}, 
R.~J.~van Weeren$^{1,}$$^{2,}$$^{3}$,
F.~Salgado$^{2}$,
L.~K.~Morabito$^{2}$,\newauthor
A.~G.~G.~M.~Tielens$^{2}$,
H.~J.~A.~Rottgering$^{2}$,
A.~Asgekar$^{1}$,
G.~J.~White$^{4,}$$^{5}$,\newauthor
A.~Alexov$^{6,}$$^{7}$
% ...
\\ \\
(affilations can be found after the references)
}
\begin{document}

\date{Accepted ????. Received ????; in original form ????}

\pagerange{\pageref{firstpage}--\pageref{lastpage}} \pubyear{0000}

\maketitle

\label{firstpage}

\begin{abstract}
We present the first detection of carbon radio recombination line absorption along the line of sight to Cygnus~A. The observations were carried out with the LOw Frequency ARray in the 33 to 57~MHz range. These low frequency radio observations provide us with a new line of sight to study the diffuse, neutral gas in our Galaxy. To our knowledge this is the first time that foreground Milky Way recombination line absorption has been observed against a bright extragalactic background source.

By stacking 48 carbon $\alpha$ lines in the observed frequency range we detect carbon absorption with a signal-to-noise ratio of about 5. The average carbon absorption has a peak optical depth of 2$\times$10$^{-4}$, a line width of 10~km~s$^{-1}$ and a velocity of +4~km~s$^{-1}$ with respect to the local standard of rest. The associated gas is found to have an electron temperature $T_{e}\sim$~110~K and density $n_{e}\sim$~0.06~cm$^{-3}$. These properties imply that the observed carbon $\alpha$ absorption likely arises in the cold neutral medium of the Orion arm of the Milky Way. Hydrogen and helium lines were not detected to a 3$\sigma$ peak optical depth limit of 1.5$\times$10$^{-4}$ for a 4~km~s$^{-1}$ channel width.

Radio recombination lines associated with Cygnus A itself were also searched for, but are not detected. We set a 3$\sigma$ upper limit of 1.5$\times$10$^{-4}$ for the peak optical depth of these lines for a 4~km~s$^{-1}$ channel width.
\end{abstract}


\section{Introduction}
Spectral lines resulting from atoms recombining with electrons in diffuse, ionised plasma are potentially important diagnostics to probe the conditions of the emitting and absorbing gas. At low quantum numbers this gives rise to the well-known optical and near-infrared recombination lines. At higher quantum numbers the energy spacing between subsequent quantum levels decreases and a recombination line transition will emit a photon at radio wavelengths. The associated lines for high quantum numbers are therefore called Radio Recombination lines (RRLs).

RRLs can be used to obtain a wealth of information on the properties of the emitting gas \citep*[e.g.][]{Go09}. Emitting in the radio domain, these lines are unbiased by dust obscuration. At low radio frequencies ($<$1~GHz) RRLs provide a straightforward method to measure the temperature, density and ionisation of the cold neutral medium \citep*[e.g.][]{Pa89}. This information can not easily be obtained by other means, such as 21~cm neutral hydrogen measurements. 

Our own Galaxy is a copious emitter of RRLs, and come in two flavours; (i) classical RRLs and (ii) diffuse RRLs. Classical RRLs are associated with the common H{\sc ii} regions \citep*{Pa67}. Here recombination lines from hydrogen, helium and carbon are seen \citep*[e.g.][]{Ko05}. These are predominantly observed at frequencies above $\sim$1~GHz as they trace the warm (T$_{e}\sim$10$^{4}$~K), high-density (n$_{e}>$100~cm$^{-3}$) gas.

Diffuse RRLs however, are associated with the low-density, cold interstellar medium \citep[e.g.][hereafter PAE89]{Ko81,Pa89}. Here only RRLs from carbon (CRRL) are  typically observed as the ionisation levels are too low to produce hydrogen and helium lines. Diffuse CRRLs are best observed at radio frequencies below 1~GHz because they arise from stimulated emission and absorption. Whereas the classical RRLs have been studied in great detail, not much is known about the gas associated with diffuse RRLs in our Galaxy. 

Low spatial resolution measurements, having beam sizes greater than 2~degrees full width at half maximum (FWHM), at 330 MHz, 76 MHz and 34.5 MHz show that on scales of degrees the CRRL emitting and absorbing gas is widespread along the plane of the inner galaxy \citep*[][hereafter EMA95]{Ka01,Er95}. Similarly, low frequency CRRLs have also been observed toward a variety of individual sources, such as for example the starforming region DR21 and the cold dark cloud L1407 \citep*{Go91a,Go91b}. Whether this CRRL gas is associated with the outershells of H{\sc ii} regions, or part of the truly diffuse cold neutral medium (CNM) is not clear from these observations. 

Recombination lines allow us to determine the temperature and density of the recombining plasma. At low frequencies these properties are best derived by studying the behaviour of the optical depth and width of the lines as a function of their quantum number \citep*[e.g.][]{Du71,Sh75,Sh76,Sa79,Wa82}. So far, such a detailed study has only been possible for the sightline towards Cassiopeia~A (Cas~A). Here it is found that the CRRLs are associated with low-density ($n_{e}\sim$0.1~cm$^{-3}$), cold ($T_{e}\sim$70~K) intervening clouds in the Perseus and Orion spiral arms \citep*[e.g. PAE89,][]{Ka98}. These properties of the intervening gas clouds are not easily deduced from other observations and CRRLs thus have the potential to provide unique information on the conditions of the CNM.

The sightline towards Cas~A is special because it shows CRRLs with optical depths that are typically about an order of magnitude brighter than those seen along other directions \citep[e.g.][EMA95,PAE89]{As13}. With the improved capabilities of new low frequency radio telescopes, in particular their broader frequency coverage and better spatial resolution, it will be possible to study the details of the RRL emitting/absorbing gas along many more sightlines and in more detail. 

Here we report on our discovery of low frequency CRRLs in absorption along the line of sight to Cygnus~A (Cyg~A) with the LOw Frequency ARray (LOFAR, \citet{Ha13}). This is an important sightline and well studied at radio wavelengths \citep*[e.g.][]{Me75,Ka96,Ca98}. Cyg~A itself is one of the brightest low frequency radio sources in the sky \citep[e.g.][McKean et al. in prep.]{La06}. This powerful FRII radio source is associated with a cD galaxy (PGC~063932) at a redshift $z=0.056$ \citep{Ow97}. 

The sightline towards Cyg~A cuts through the Milky Way at a galactic longitude $l$~=~76.19${\degree}$ and latitude $b$~=~+5.76${\degree}$. H{\sc i} 21~cm line observations show both emission and absorption in the range +30 to $-$120~km~s$^{-1}$ \citep*[e.g.][]{Ca98}. The strongest H{\sc i} absorption features are found around +5 and $-$85~km~s$^{-1}$. These are associated with the CNM in Orion (+5~km~s$^{-1}$) and the outer spiral arms ($-$85~km~s$^{-1}$). The H{\sc i} observations of Cyg~A also show absorption from the warm neutral medium (WNM) at velocities of $-$40 and $-$70~km~s$^{-1}$ \citep{Ca98,Me75}.

This paper is structured as follows. Our observations are summarised in Section~2 and the results are presented in Section~3. We discuss these results in Section~4 and present our conclusions in Section~5.

\section[]{Observations and Reduction}\label{s_obs_red}
We observed Cyg~A with the LOFAR Low-band antenna (LBA) on 2011 December 27 from 10:00 to 20:30 UTC. The observations were taken during the commissioning phase of the array. We obtained complete frequency coverage between 33 and 57~MHz, although a few sub-bands were corrupted due to issues with the LOFAR offline storage system, see Table~\ref{t_obs}. The bandwidth was divided in 122 sub-bands, each 0.1953 MHz wide. The LBA\_OUTER configuration was used for the LBA stations and all four linear correlation products were recorded (XX, XY, YX, YY). Each sub-band was subdivided into 512 frequency channels. The visibility integration time was 2~s. We used 9 remote and 22 core stations for this observation, giving baselines between 90~m and 85~km.

As a first step in the data processing, we flagged radio frequency interference (RFI) with the AOFlagger \citep{Of10}. Typically, a few percent of the data were flagged due to RFI, which is consistent with a study of the LOFAR RFI environment conducted by \citet{Of13}. The data were calibrated with the BlackBoard Selfcal (BBS) software system \citep{Pa09}. We used a high-resolution 10~arcsec clean component model of Cyg~A for calibration. This model was obtained from previous LOFAR observations around 70~MHz (McKean et al. in prep).

Before calibrating, we made a copy of the dataset, averaging each sub-band down from 512 to 1 channel. We then obtained gain solutions for all four correlations with BBS on a 4~s timescale. We assumed that the source is unpolarized over the observed frequency range. The found gain solutions were then applied to the 512~frequency channel data and a final round of flagging was carried out with the AOFlagger. 

Channel cubes were made with CASA\footnote{http://casa.nrao.edu/}, imaging and cleaning each channel individually. The first 25 and last 25 channels of each sub-band were ignored as they are too noisy. Only baselines with a $uv$-distance shorter than 7~km were selected. We chose Briggs weighting \citep{Br95} with a robust value of 0.5 to create images with a resolution ranging between 174$\times$196~arcsec$^{2}$ and 267$\times$336~arcsec$^{2}$. This low resolution was chosen to speed up the imaging proces. LOFAR does currently not support Doppler tracking and as such we have Doppler corrected the data during the imaging process. This introduces a small uncertainty ($<$1~km~s$^{-1}$) in the derived line widths and velocities, but does not affect our analysis of the data. We then convolved all images from all sub-bands to a common resolution of 300$\times$350~arcsec$^{2}$. At this resolution Cyg~A is not resolved.

Our final cubes for each sub-band image a region with an area of 1$\times$1~degree$^{2}$, with Cyg~A at the centre. From these cubes we extract spatially integrated on-source spectra from a 15$\times$15~arcmin$^{2}$ aperture centred on Cyg~A. To estimate the continuum in these on-source spectra we fit a 4th order polynomial. We then divide the spectra by the fitted continuum (and subtract by 1) to derive the final on-source sub-band spectra in units of optical depth, i.e. $\tau_{\nu}$~=~(F$_{\nu,obs}$/F$_{\nu,cont}$)~-~1. Here F$_{\nu,obs}$ is the observed flux and F$_{\nu,cont}$ is the continuum flux. Out of a total of 122 sub-bands seven were removed due to data corruption. For the remaining 115 sub-bands spectra were extracted and visually inspected. Subsequently, ten additional sub-bands were removed due to strong RFI in the data, leaving 105 sub-bands for further analysis.

The on-source spectra for each sub-band were investigated for the presence of RRLs. Their peak optical depth limits, that is the root-mean-square (RMS) per 0.4~kHz channel divided by the continuum level, varied from about 3$\times$10$^{-4}$ at 33 MHz to 8$\times$10$^{-4}$ at 57 MHz. At these levels RRLs were not detected in the individual sub-bands.

We proceeded by stacking the on-source spectra. We only included spectra where the expected line position was at least 70 channels away from the edge of the subband to avoid errors due to the bandpass rolloff. Stacking was done in velocity space using a 2~km~s$^{-1}$ channel spacing corresponding to our highest spectroscopic resolution at 57~MHz. The final stacked spectrum is rebinned to 4~km~s$^{-1}$ which corresponds to our lowest spectroscopic resolution at 33~MHz. 

Stacked spectra were produced for RRL $\alpha$ transition lines from hydrogen, helium and carbon at redshifts corresponding to (i) the Milky Way foreground ($z=0$) and (ii) Cyg~A ($z=0.056$). In each case about 50 lines were stacked. The noise in units of root-mean-square (RMS) per channel divided by the continuum level is found to decrease from about 4$\times$10$^{-4}$ for the on-source spectra from individual sub-bands to about 0.5$\times$10$^{-4}$ for the final stacked spectra. The decrease in the noise thus roughly scales with the square root of the number of stacked spectra. This is consistent with Gaussian spectral noise in the on-source spectra from individual sub-bands. 

\section[]{Results}\label{s_results}
We searched the stacked spectra for RRLs associated with both the Milky Way foreground ($z=0$) and with Cyg~A ($z=0.056$). For the Milky Way foreground stacked spectra we find an absorption feature with a signal-to-noise ratio (SNR) of 5. This absorption feature is found at a velocity of $+$4~km~s$^{-1}$, relative to the local standard of rest (LSR), if the stack is centred on the carbon $\alpha$ transitions, see Fig.~\ref{f_results}. Similarly the absorption feature is found at the expected velocity of $-$30~km~s$^{-1}$ and $-$150~km~s$^{-1}$ when centring on the helium and hydrogen $\alpha$ transitions, respectively, see Fig.~\ref{f2_results}. We believe that the observed absorption feature is due to carbon and we present our arguments for this in Section~\ref{s_discuss_s1}.

\subsection{Detected Carbon RRLs at $z=0$}
\begin{figure*}
\mbox{
    \includegraphics[width=0.35\textwidth, angle=90]{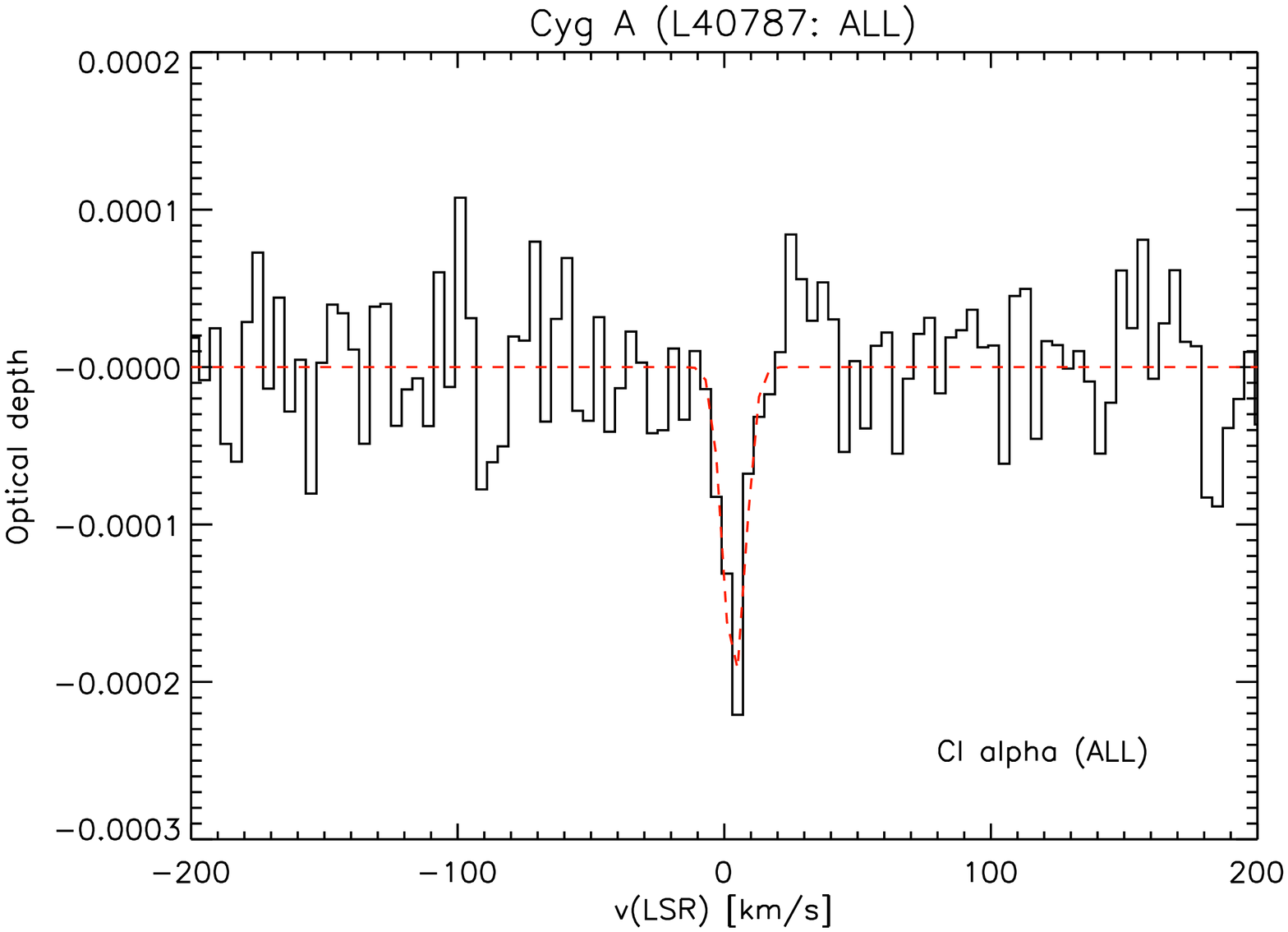}
    \includegraphics[width=0.35\textwidth, angle=90]{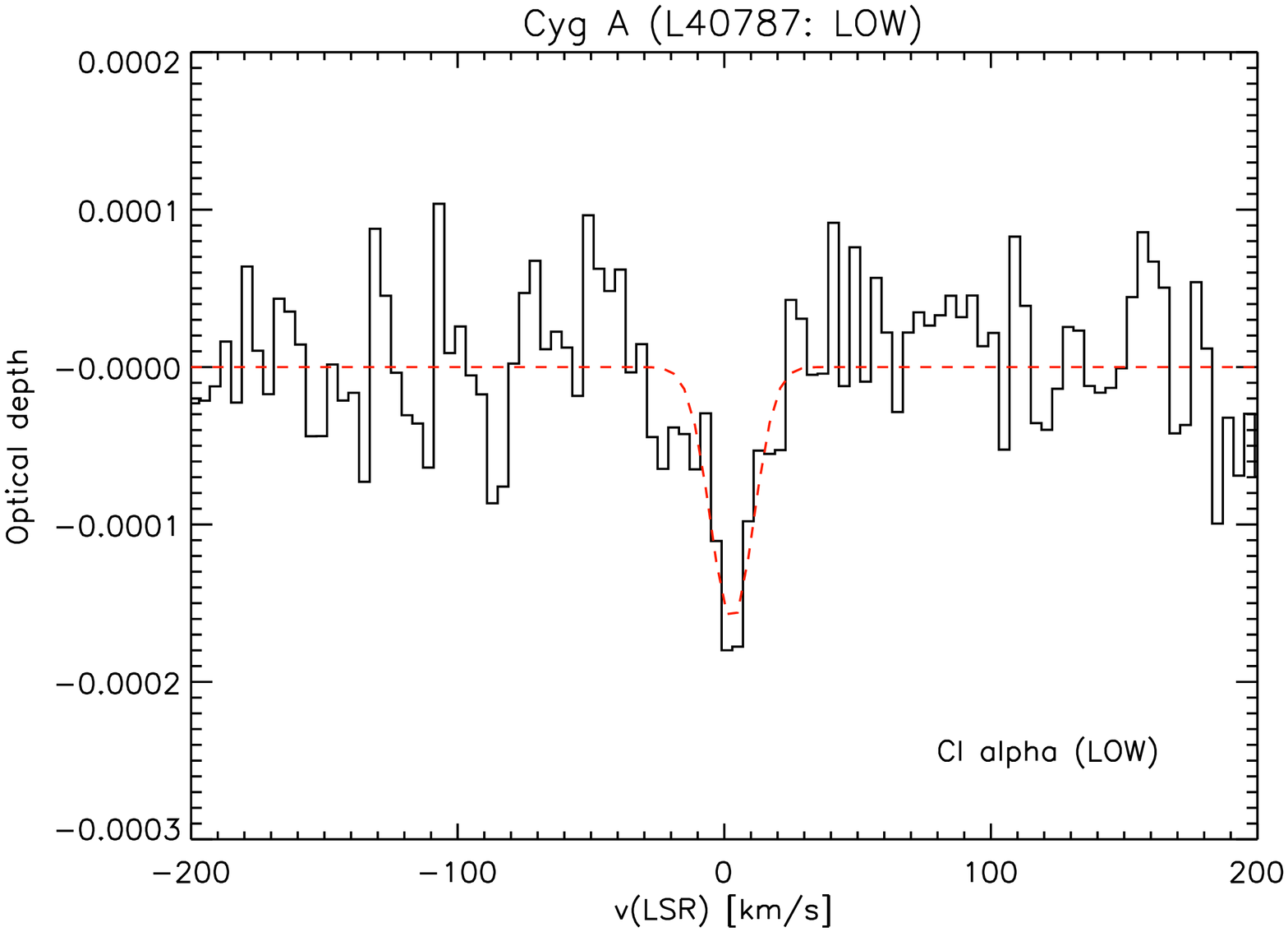}
}
\mbox{
    \includegraphics[width=0.35\textwidth, angle=90]{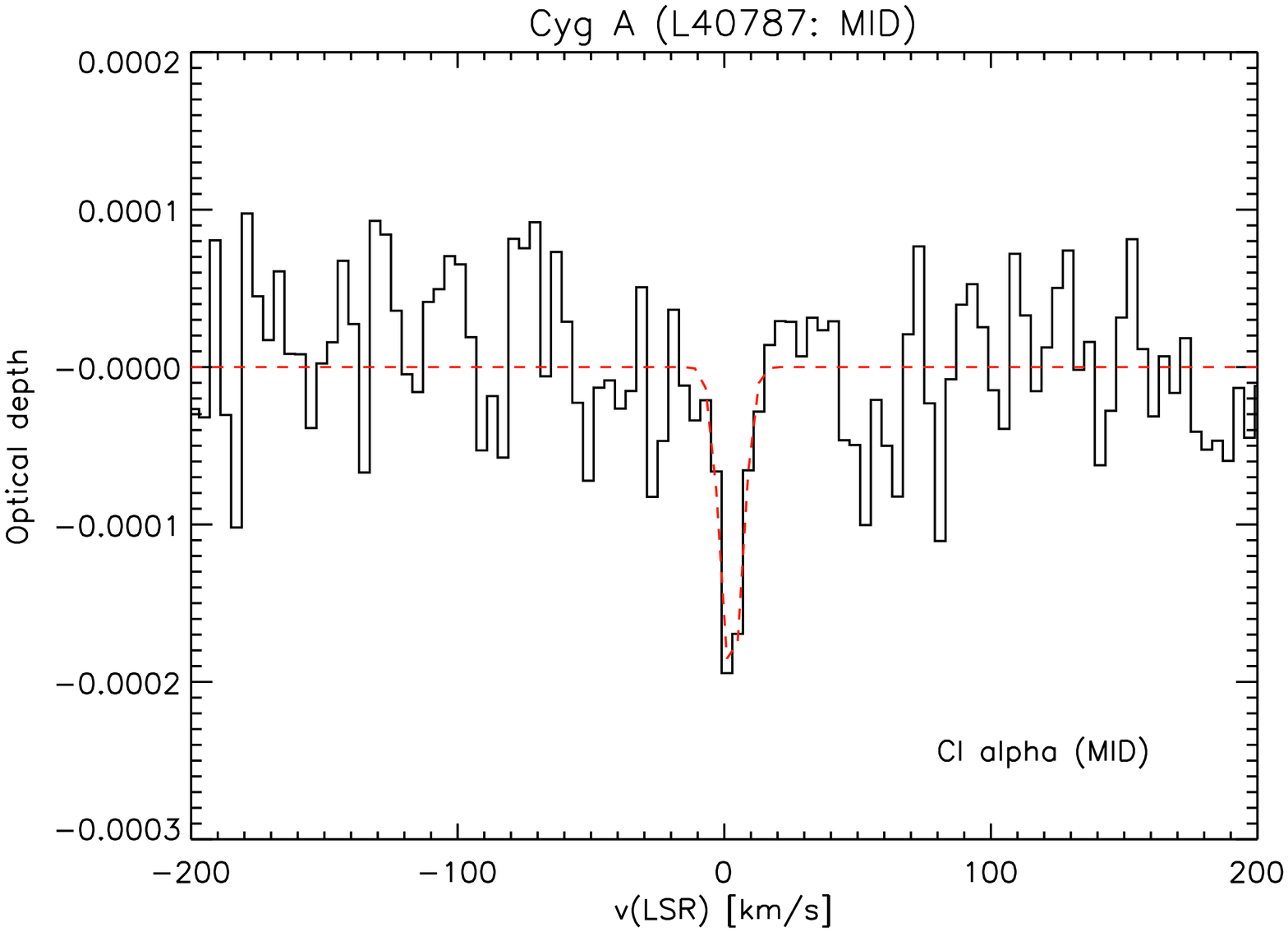}
    \includegraphics[width=0.35\textwidth, angle=90]{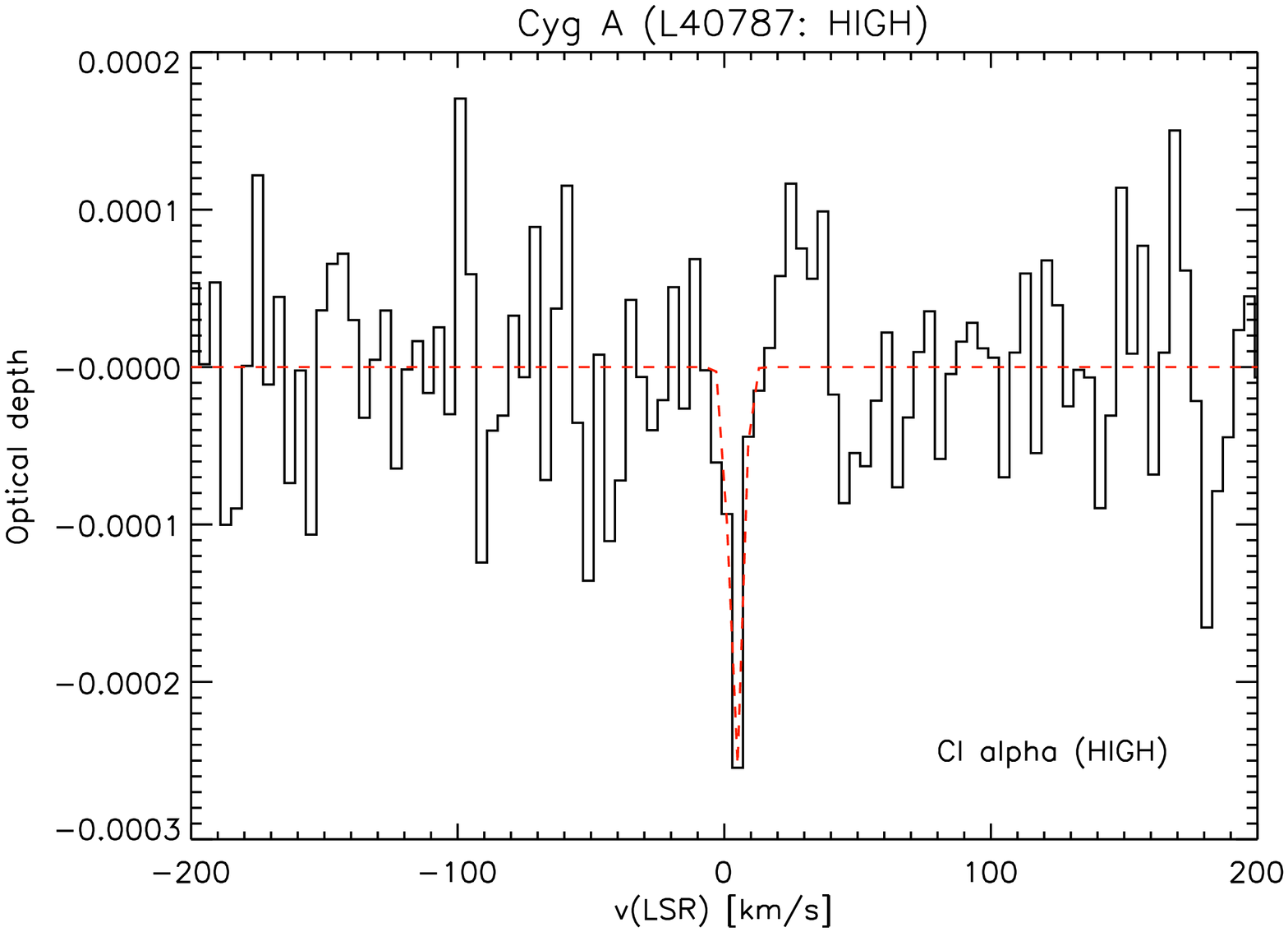}
}
  \vspace{0.5cm}
  \caption{Foreground stacked carbon $\alpha$ spectrum along the line of sight to Cyg~A. The data are stacked assuming a redshift $z=0$ for the lines and displayed using a resolution of 4 km~s$^{-1}$ per channel. (\textit{Top-Left}) Stacked spectrum of all 48 carbon $\alpha$ lines in the range 33 to 57 MHz. (\textit{Top-Right}) Stacked spectrum of the 21 lowest frequency carbon $\alpha$ lines. (\textit{Bottom-Left}) Stacked spectrum of 23 carbon $\alpha$ lines in the middle of the observed range. (\textit{Bottom-Right}) Stacked spectrum of the 27 highest frequency carbon $\alpha$ lines. The black line shows the data and the red line shows the Gaussian fit. The line properties derived from these fits are given in Table \ref{t_results}.}\label{f_results}
\end{figure*}

For carbon $\alpha$ transition RRLs from the Milky Way foreground, a total of 48 on-source sub-band spectra are stacked. The resulting stacked spectrum, in units of optical depth, is shown in the top-left corner of Fig.~\ref{f_results}. Henceforth we will refer to this stack as ALL. This carbon $\alpha$ spectrum from the ALL stack shows a RRL absorption feature with a peak optical depth of about 2$\times$10$^{-4}$. A single Gaussian fit to this feature gives a full width at half maximum (FWHM) line width of 10~km~s$^{-1}$ and a central velocity of $+$4~km~s$^{-1}$ relative to the LSR, see Table \ref{t_results}.

All of the sub-band spectra entering the stack were carefully inspected by eye. We then performed a 'Jacknife minus one' procedure to investigate whether the observed carbon absorption could have been caused by a rogue feature from any single spectrum. This method consists of leaving out one sub-band spectrum at a time and recomputing the stack. We did this for all 48 sub-band spectra, thus creating 47 stacks of 47 sub-band spectra. For each of these stacks, the absorption line and noise properties are found to be consistent with the full 48 sub-band stack.

We then split our data into three sub-stacks; LOW, MID and HIGH (Table~\ref{t_results}). The LOW sub-stack contains the 21 lowest frequency lines and the HIGH sub-stack contains the 27 highest frequency lines. These two stacks do not overlap, and therefore represent two independent measurements. The MID sub-stack contains 23 lines from the middle portion of the observed frequency range. This stack overlaps with both the LOW and HIGH sub-stack. These three stacks were made to check for (i) variation/changes in the line profile with frequency, and (ii) the possibility of having rogue features from more than one spectrum being responsible for the observed RRL absorption. In all sub-stacks carbon $\alpha$ RRL absorption is detected, see Fig.~\ref{f_results}. 

Finally, we investigated the off-source spectral properties by excluding the central 15$\times$15~arcmin$^{2}$ aperture containing Cyg~A from the data, but including the rest of the 1$\times$1~deg$^{2}$ field of view. The stacked off-source spectra did not show any spectral features that could have contaminated the above results. We therefore conclude that observed RRL absorption is real and associated with the line-of-sight to Cyg~A. The SNR of the integrated optical depth is about 5 for the ALL stack and between 3 and 4 for the sub-stacks.

We also searched for carbon $\beta$ transition RRLs from the Milky Way foreground but do not detect these. This non-detection is not surprising as the carbon $\beta$ lines in our Galaxy are typically a factor of about three weaker than the $\alpha$ lines (e.g. EMA95, PAE89) and therefore beyond the sensitivity limit of our observation. We set a 3$\sigma$ upper limit of 1.5$\times$10$^{-4}$ for the peak optical depth of these lines for a 4~km~s$^{-1}$ channel width.

\subsection{Upper limits on H and He RRLs at $z=0$}
\begin{figure*}
\mbox{
    \includegraphics[width=0.35\textwidth, angle=90]{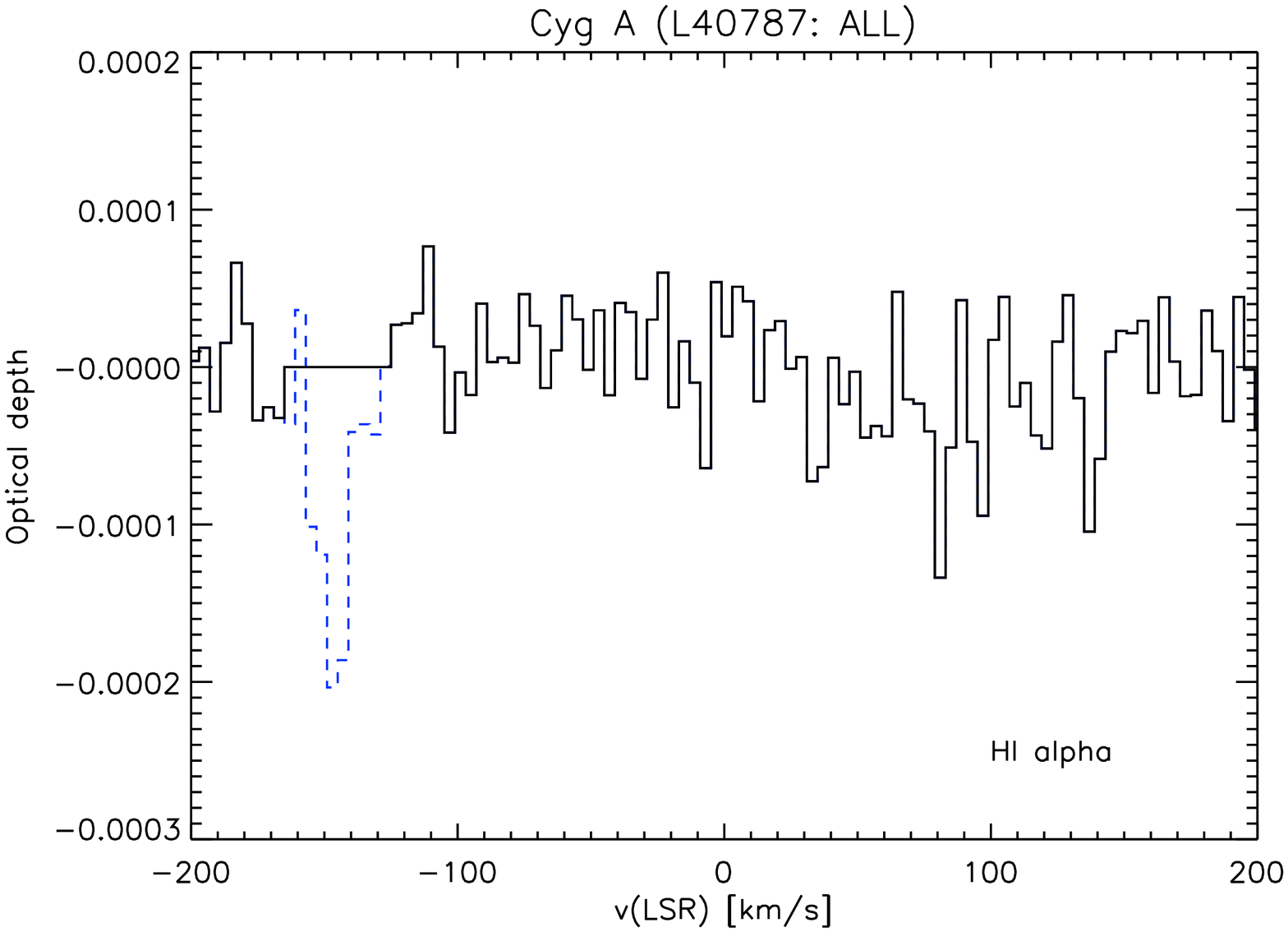}
    \includegraphics[width=0.35\textwidth, angle=90]{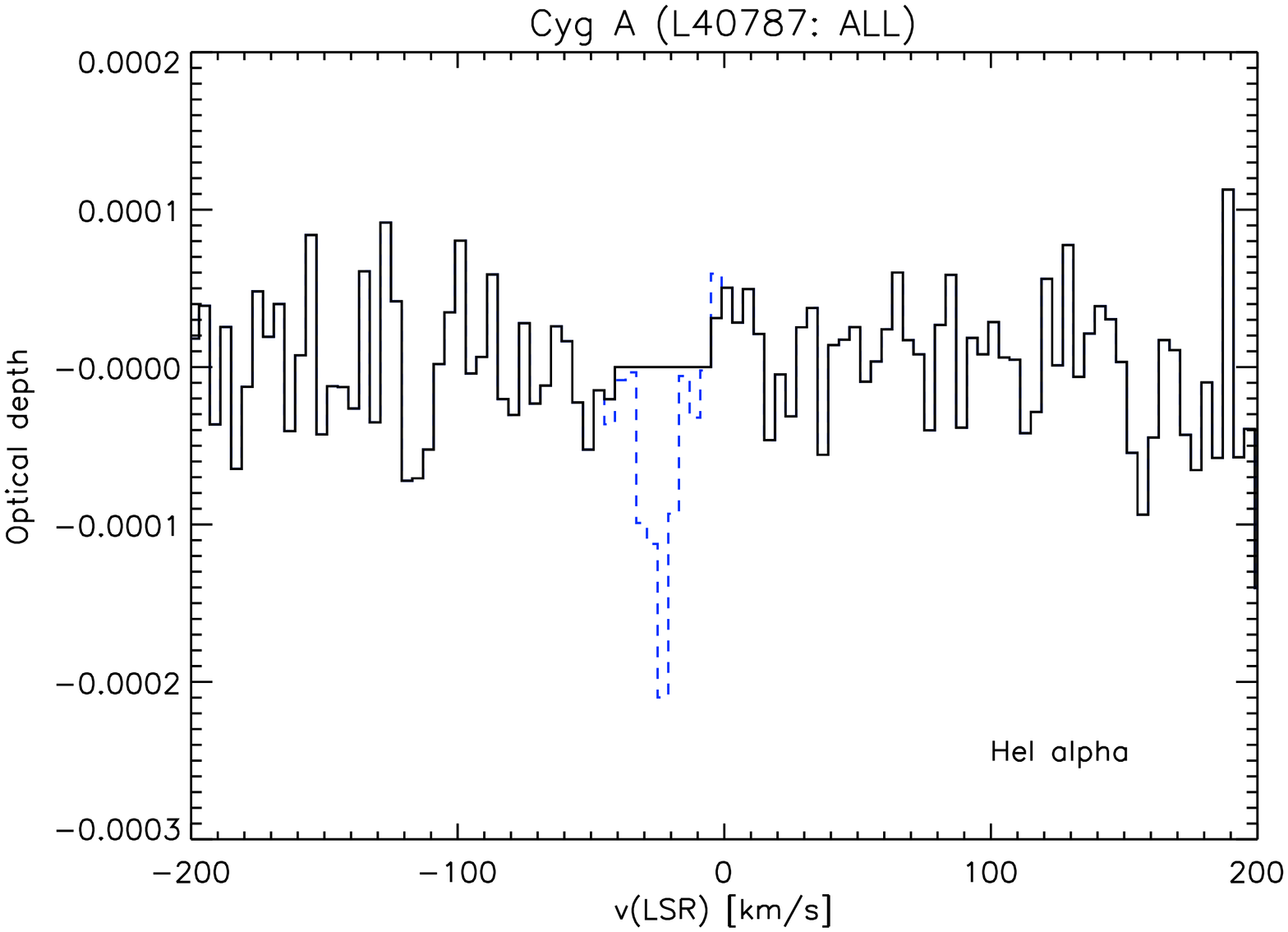}
}
  \vspace{0.5cm}
  \caption{Foreground stacked hydrogen $\alpha$ and helium $\alpha$ spectra along the line-of-sight to Cyg~A. The data are stacked assuming a redshift $z=0$ for the lines and displayed using a resolution of 4 km~s$^{-1}$ per channel. (\textit{Left}) Stacked spectrum of all 49 hydrogen $\alpha$ lines in the range 33-57 MHz. The region around $-$150 km~s$^{-1}$ corresponds to carbon $\alpha$ line absorption and was blanked in all spectra prior to stacking. (\textit{Right}) Stacked spectrum of all 48 helium $\alpha$ lines in the range 33-57 MHz. The region around $-$30 km~s$^{-1}$ corresponds to carbon $\alpha$ line absorption and was blanked in all spectra prior to stacking. For reference we have overplotted, using the blue dashed line, the stacked hydrogen, helium spectra where the carbon $\alpha$ line absorption was not blanked apriori. The upper limits derived from the hydrogen, helium spectra are given in Table \ref{t2_results}.}\label{f2_results}
\end{figure*}

Stacked spectra centred on the rest frequency of the hydrogen and helium RRLs for the Milky Way foreground ($z=0$) were also produced, see Fig.~\ref{f2_results}. Prior to stacking the spectra we now first blank the carbon $\alpha$ absorption feature in each spectrum. If this is not done the absorption feature seen in the stacked carbon spectrum shows up at about $-$150 km~s$^{-1}$ in the stacked hydrogen spectrum and at about $-$30 km~s$^{-1}$ in the stacked helium spectrum. To illustrate this we show the unblanked hydrogen and helium stacked spectra with the blue dashed line in Fig.~\ref{f2_results}. 

We expect the absorption feature seen in the carbon $\alpha$ stack to also show up in the hydrogen and helium spectra. This is because the RRL rest frequencies for these elements have a constant offset in velocity space. At 45 MHz ($n=526$) this corresponds to $-$22.5~kHz ($-$150~km~s$^{-1}$) for hydrogen relative to carbon and $-$4.1~kHz ($-$27~km~s$^{-1}$) for helium relative to carbon. The relatively small offset in velocity means that in principle the absorption feature could come from any of these three elements. However, as we will discuss below the observed feature is most likely associated with carbon.

The 3$\sigma$ upper limits on the peak optical depth $\tau_{\rm{peak}}$ that we obtain from the $z=0$ stacked hydrogen and helium spectra are about 1.5$\times$10$^{-4}$ for a channel width of 4~km~s$^{-1}$, see Table~\ref{t2_results}.

\subsection{Upper limits on RRLs at $z=0.056$}
\begin{figure*}
\mbox{
    \includegraphics[width=0.24\textwidth, angle=90]{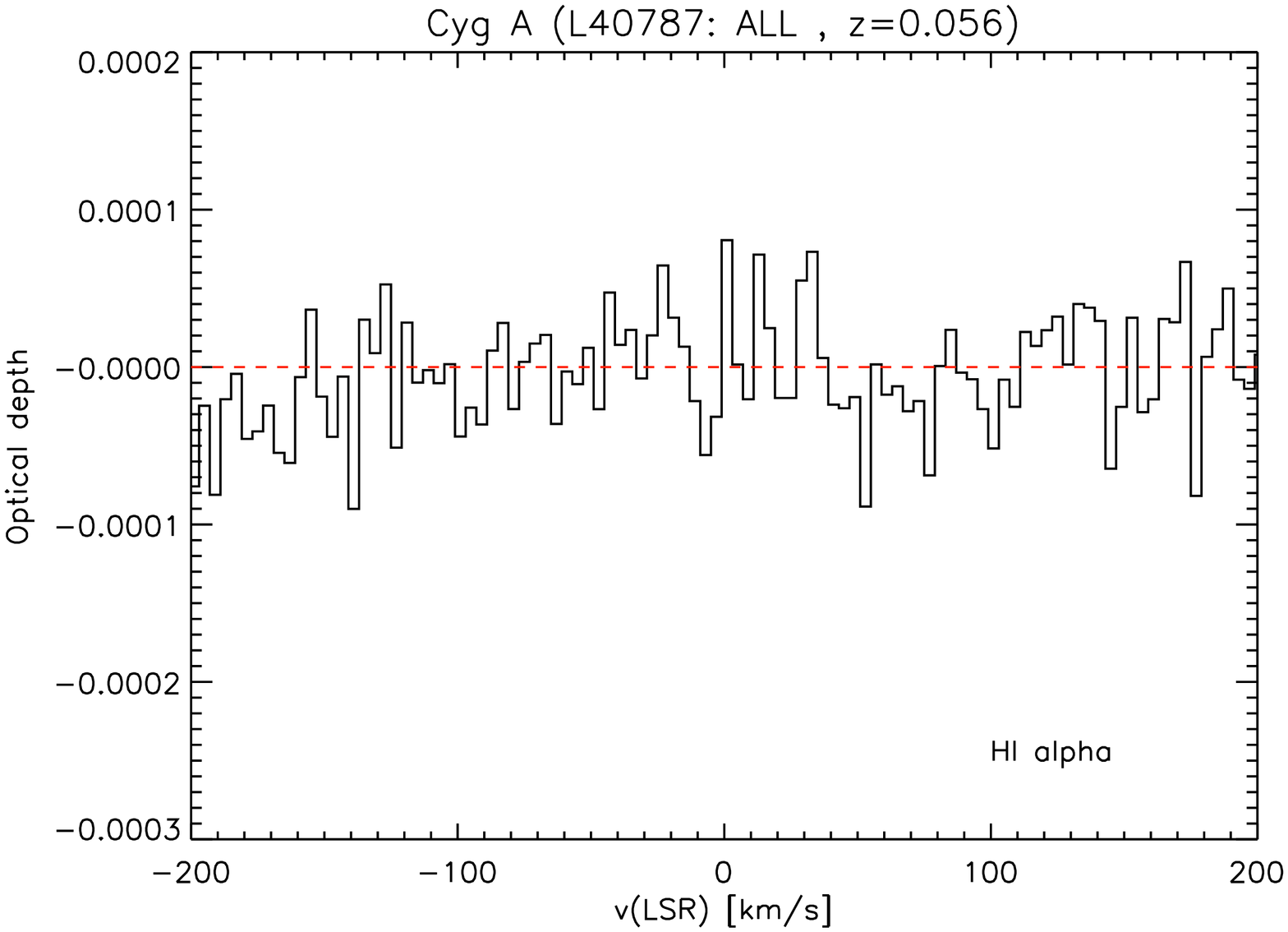}
    \includegraphics[width=0.24\textwidth, angle=90]{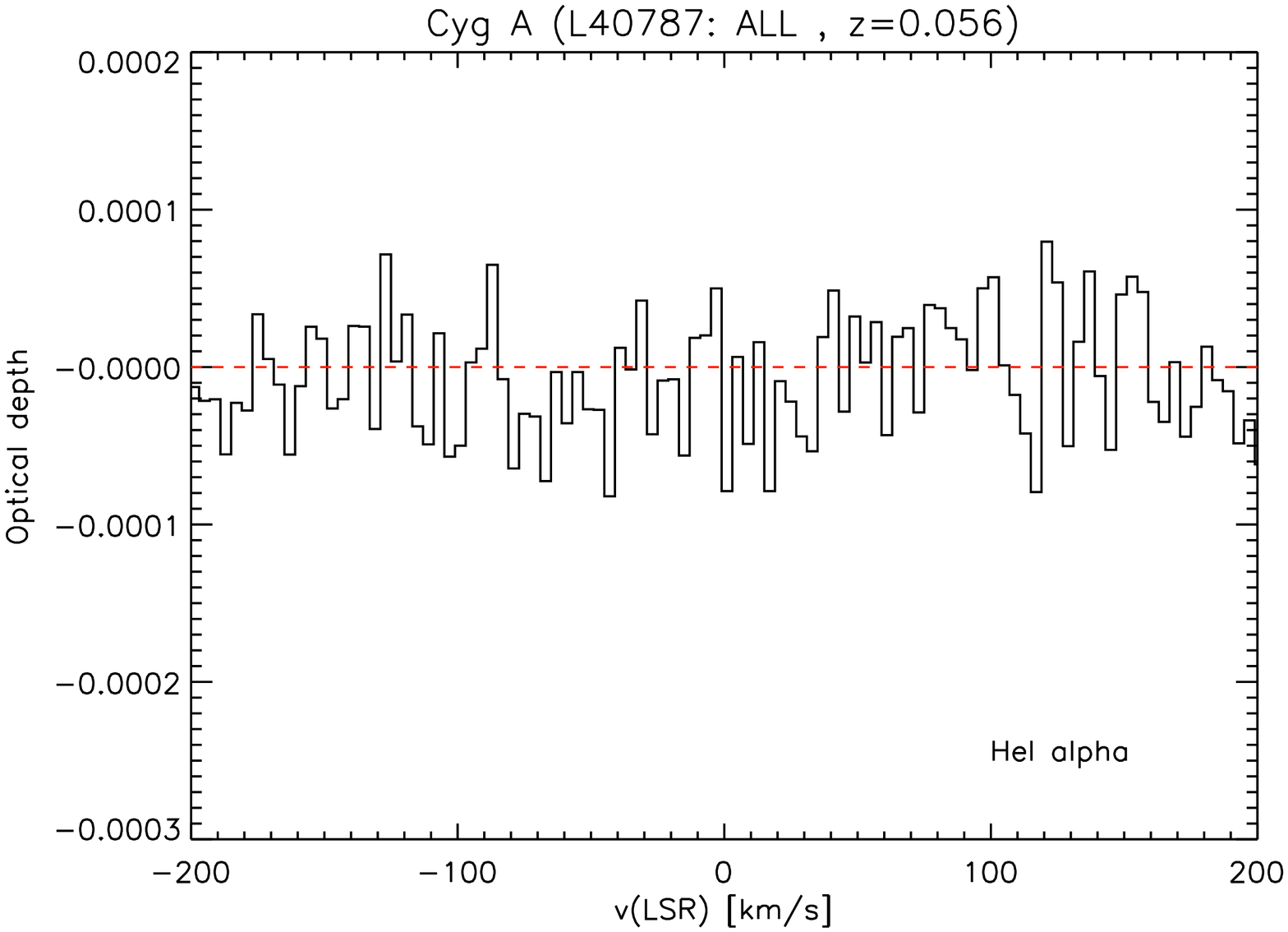}
    \includegraphics[width=0.24\textwidth, angle=90]{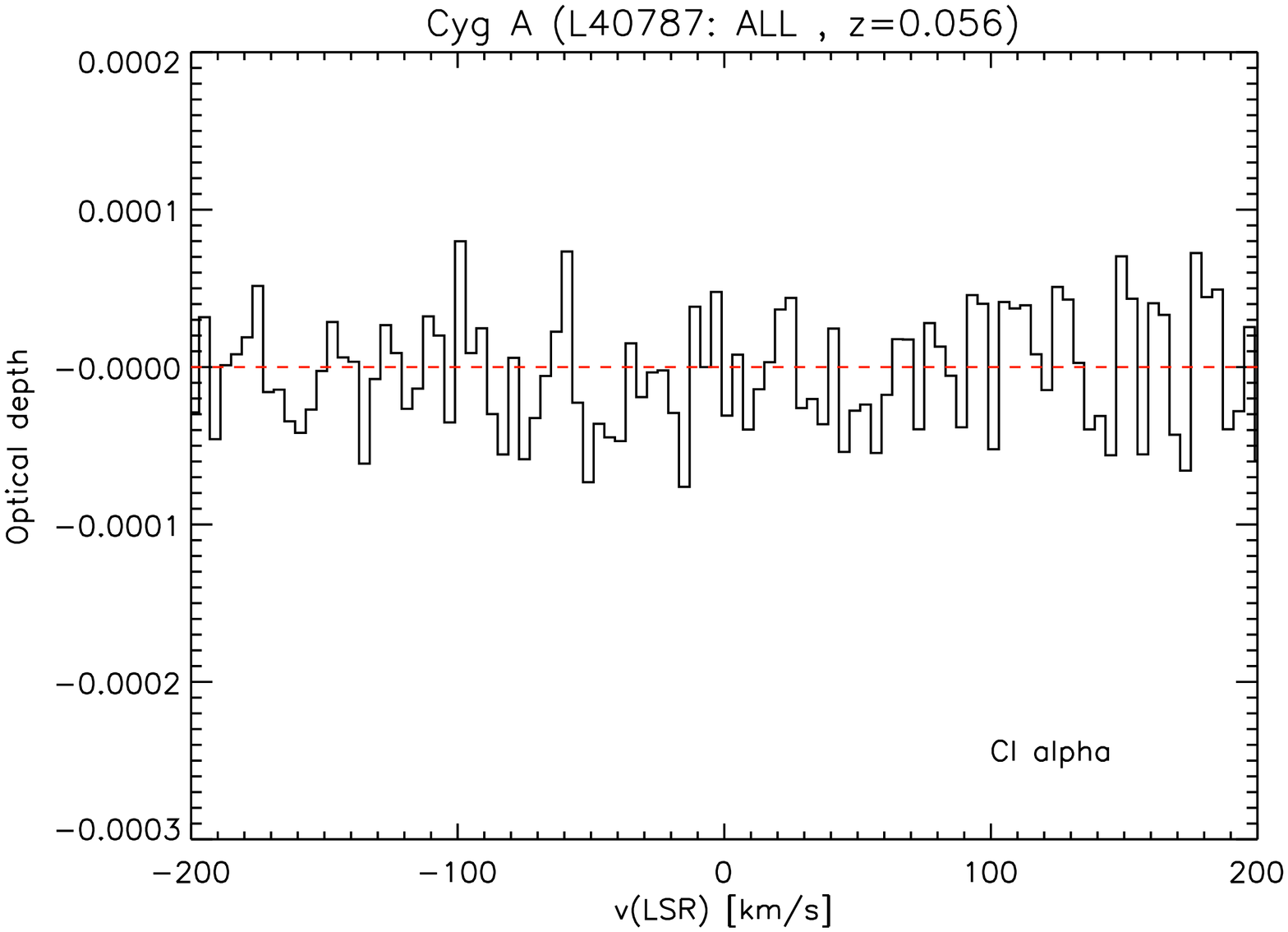}
}
  \vspace{0.5cm}
  \caption{Stacked hydrogen $\alpha$, helium $\alpha$ and carbon $\alpha$ spectra for Cyg~A. The data is stacked assuming at redshift $z=0.056$ for the lines and displayed using a resolution of 4 km~s$^{-1}$ per channel. (\textit{Left}) Stacked spectrum of all 55 hydrogen $\alpha$ lines in the range 33 to 57 MHz. (\textit{Middle}) Stacked spectrum of all 50 helium $\alpha$ lines in the range 33 to 57 MHz. (\textit{Right}) Stacked spectrum of all 47 carbon $\alpha$ lines in the range 33 to 57 MHz. The upper limits derived from these spectra are given in Table \ref{t2_results}.}\label{f3_results}
\end{figure*}

In the same manner as described above for the Milky Way foreground gas, we also produced stacked hydrogen, helium and carbon RRL spectra in the redshift range 0.046 to 0.066, i.e. $\pm$3000~km~s$^{-1}$ around the optical emission line redshift 0.056 for Cyg~A. The resulting spectra show no sign of RRLs associated with this source. As an example we show the stacked spectra at the redshift $z=0.056$ in Fig.~\ref{f3_results}.

The 3$\sigma$ upper limits on the peak optical depth $\tau_{\rm{peak}}$ that we obtain from the $z=0.056$ stacked hydrogen, helium and carbon spectra are about 1.3$\times$10$^{-4}$ for a channel width of 4~km~s$^{-1}$, see Table~\ref{t2_results}.

\section[]{Discussion}\label{s_discuss}

\subsection{A Carbon origin for the RRL absorption}\label{s_discuss_s1}
The foreground RRL absorption feature observed here against Cyg~A ($l$~=~76.19$\degree$ , $b$~=~+5.76$\degree$) is very narrow and close to zero velocity. A similar RRL feature is seen towards Cas~A ($l$~=~111.74$\degree$, $b$~=~-2.14$\degree$) by, for example, PAE89. These authors show that the RRL absorption along the line of sight to Cas~A consists of three velocity components at $-$48, $-$37 and 0~km~s$^{-1}$. The first two components have an optical depth which is about a factor of 10 larger than the component at zero velocity. 

PAE89 show that all three RRL components are carbon RRLs. The $-$48 and $-$37~km~s$^{-1}$ components are related to the Perseus spiral arm, whereas the zero velocity component is related to the Orion spiral arm. The similarity between the zero velocity component towards Cyg~A and Cas~A, in both line width and peak optical depth, indicates that the absorption seen by us against Cyg~A is probably coming from carbon RRLs and likely originates from the CNM in the Orion spiral arm of the Milky Way. 

If the observed absorption feature were instead due to helium at a velocity of $-$30~km~s$^{-1}$ then we would expect to not only observe helium, but also hydrogen and carbon absorption. This is because (i) hydrogen is much more abundant than helium, and (ii) the high ionisation potential of helium implies that hydrogen and carbon would also be ionised.

\begin{figure*}
\mbox{
    \includegraphics[width=0.35\textwidth, angle=90]{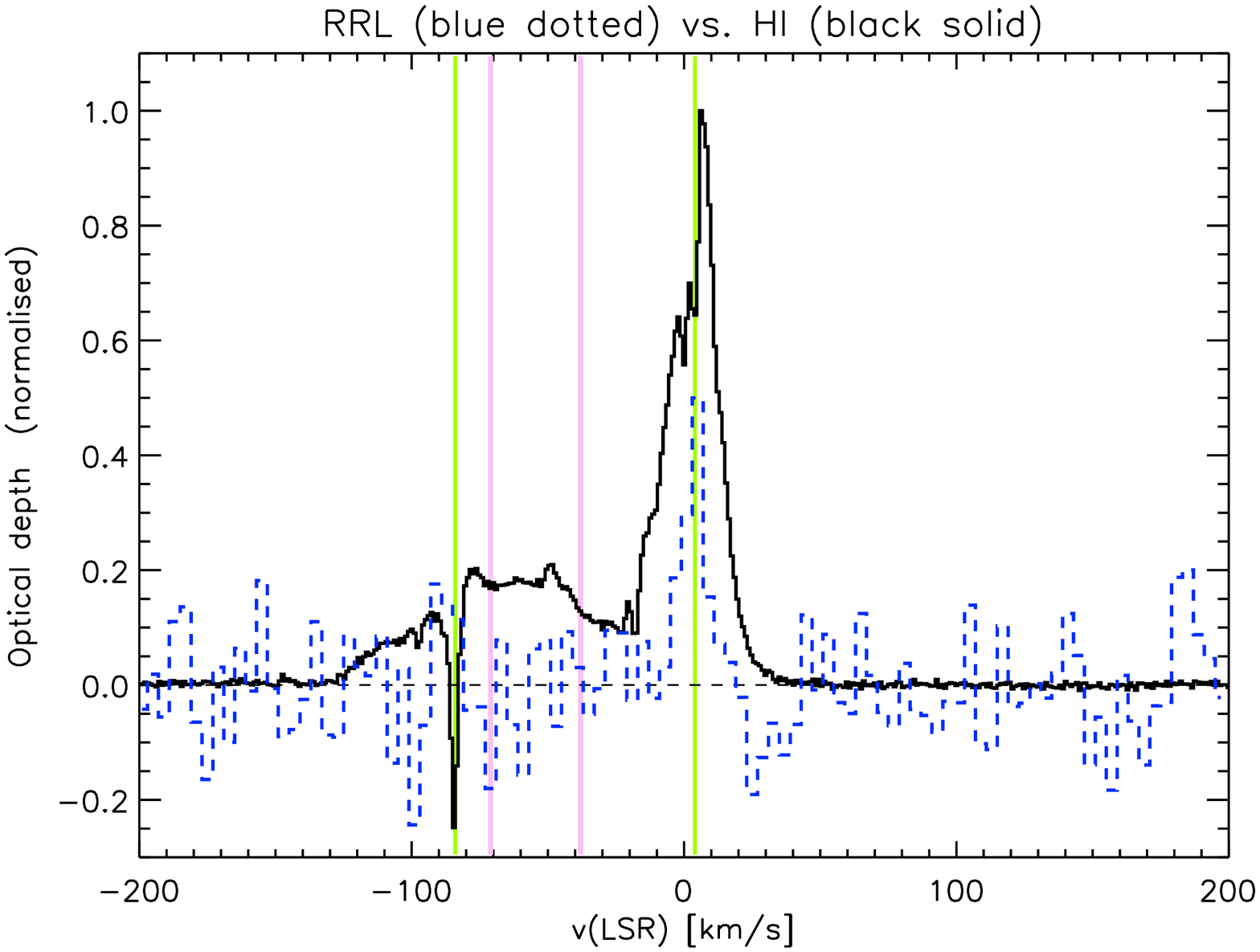}
    \includegraphics[width=0.35\textwidth, angle=90]{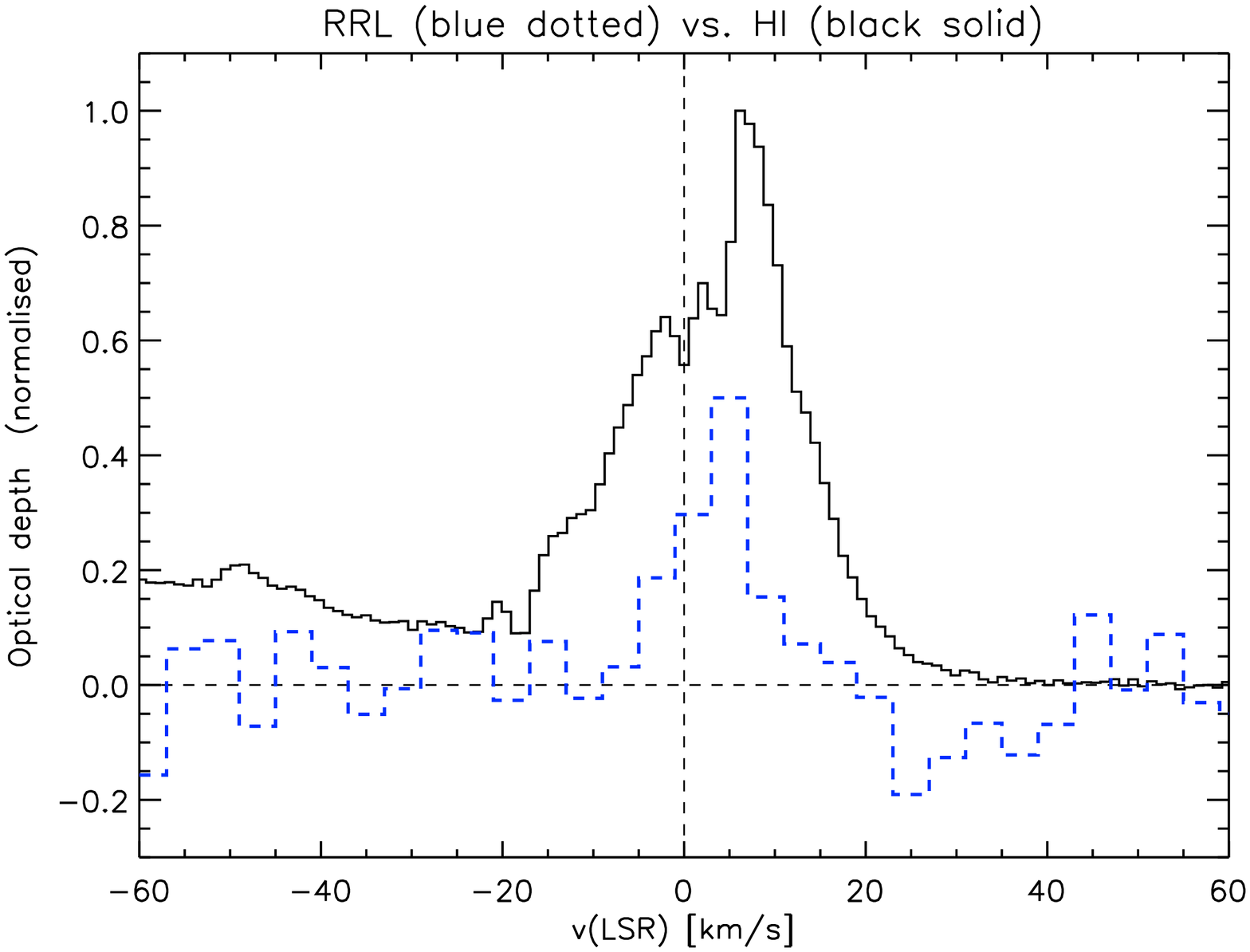}
}
  \vspace{0.5cm}
  \caption{Comparison of the foreground ($z=0$) stacked carbon $\alpha$ absorption with the H{\sc i} emission along the line of sight to Cyg~A. The H{\sc i} emission, from the LAB survey (Kalberla et al. 2005), is given by the black line. The inverted carbon $\alpha$ absorption (this work) is shown by the dashed blue line. For clarity the H{\sc i} emission has been normalised to 1.0 and the inverted RRL absorption has been normalised to 0.5. \textit{(Left)} H{\sc i} and RRL in the velocity range $\pm$200~km~s$^{-1}$. The vertical green (CNM) and purple (WNM) lines show the positions of the strongest H{\sc i} 21~cm components observed in absorption towards Cyg~A (Carilli et al. 1998). \textit{(Right)} The central $\pm$60~km~s$^{-1}$ range.}\label{f_compare}
\end{figure*}

We believe that it is unlikely that we are observing hydrogen RRL absorption at a velocity of $-$150~km~s$^{-1}$. This is because the bulk of the H{\sc i} 21~cm emission, as observed in the Leiden-Argentine-Bonn survey \citep{Ka05}, along the line of sight to Cyg~A is found at zero velocity, see Fig.~\ref{f_compare}. 

The H{\sc i} 21~cm emission line profile from \citet{Ka05} is much broader than what we observe for the RRL absorption. This is likely due to the difference in beam size for the two observations. The 21~cm observation has a FWHM beam size of about 40~arcmin which is significantly larger than our RRL observation. It therefore samples a much greater area, having a range of gas velocities that contribute to the broadening of the observed H{\sc i} 21~cm line profile.

H{\sc i} absorption observed against Cyg~A, by \citet{Ca98} and \citet{Me75}, shows much narrower H{\sc i} absorption line profiles. These authors find several absorbing components in the range +30 to $-$120~km~s$^{-1}$. The strongest component is found at zero velocity and this component has a width similar to our RRL absorption feature. The H{\sc i} absorption spectrum shows no significant absorption at either $-$30 or $-$150 km~s$^{-1}$. We therefore conclude that the RRL absorption observed by us towards Cyg~A is most likely due to carbon. 

Comparing the RRL and H{\sc i} absorption furthermore shows that the RRL absorbing gas is likely associated with the CNM in the Orion arm \citep{Ca98,Me75}. None of the other H{\sc i} absorbing components, from either the CNM or WNM, are detected in our RRL spectrum, see Fig.~\ref{f_compare}. This is not surprising given the low sensitivity of our stacked spectrum. There may be a hint of the outer spiral arm ($-$85~km~s$^{-1}$) component in the CRRL spectrum, but more sensitive observations are required to confirm this. Higher spatial resolution imaging of our data may in the future also help to investigate the presence of this component. This is because the H{\sc i} optical depth of the $-$85~km~s$^{-1}$ component is observed to vary strongly in strength across the radio source \citep{Ca98}.

\subsection{Gas properties from Carbon RRLs}\label{s_modeling_ab}
Carbon RRLs are not in local thermal equilibrium (LTE). Therefore to obtain physical parameters for the CRRL emitting gas one needs to perform detailed modelling of the transitions \citep*[e.g.][]{Du69,Sh75}. The two observables that constrain the properties of the associated gas are the integrated optical depth and the width of the CRRL line.

The integrated optical depth, in units of s$^{-1}$, can be written as \citep*{Sh75},

\begin{eqnarray}
 \int \tau_{Cn}~d\nu &=& 2.046 \times 10^{6}~T_{e}^{-5/2}~e^{\chi_{n}}~\rm{EM}_{C}~(b_{n}\beta_{n})_{C}
\end{eqnarray}

Here $n$ is the quantum number, $T_{e}$ is the electron temperature, $\chi_{n}=1.58\times$10$^{5}/(n^{2}T_{e})$ and $EM_{C}=n_{e}n_{C+}L$ [cm$^{-6}$~pc] is the emission measure for carbon, with $L$ the path length. $b_{n}$, $\beta_{n}$ are the departure coefficients from LTE and were calculated following \citet{Sh75} and \citet{Wa82} using a modified version of the \citet*{Br77} code that takes the dielectronic-like recombination of carbon into account (Salgado et al. in prep.). These departure coefficients depend on $T_{e}$ and the electron density $n_{e}$.

The current dataset only provides a rough constraint on these quantities. Based on a Chi-Squared minimization, using a fine grid of models with T$_{e}$ between 10 and 1000~K and n$_{e}$ between 0.001 and 1.0~cm$^{-3}$, we find that the data are best described by an electron temperature $T_{e}\approx 110$~K, electron density $n_{e}\approx 0.06$~cm$^{-3}$ and $EM_{C}\approx 0.001$~cm$^{-6}$~pc. The gas traced by the CRRL absorption is thus found to be consistent with the cold neutral medium. However, the large uncertainties and limited frequency coverage of the measurements imply that we can only constrain $T_{e}$ to be in the range 50 to 500~K and $n_{e}$ to be in the range 0.005 to 0.07~cm$^{-3}$. These ranges for $T_{e}$ and $n_{e}$ correspond to a 3$\sigma$ confidence limit.

The CRRL line width can be broadened by (i) pressure broadening $\Delta$v$_{P}$, (ii) radiation broadening $\Delta$v$_{R}$ and (iii) multiple velocity components along the line of sight \citep{Sh75}. Although we observe a weak assymmetry in the CRRL line profile (Fig.~\ref{f_compare}), also seen in H{\sc i} absorption \citep{Ca98}, the quality of our spectrum is insufficient to determine whether more than one velocity component exists. Hence, we will only consider the line as a single component. 

We observe that the CRRL line width increases with increasing quantum number $n$, as expected from pressure and radiation broadening (Table~\ref{t_results}). In units of km~s$^{-1}$, these broadening terms can be written as \citep{Sh75},

\begin{eqnarray}
 \Delta V_{P,n} &=& 2 \times 10^{-8}~e^{-26/T_{e}^{1/3}}~\frac{n_{e}n^{5.2}}{T_{e}^{1.5}}~\frac{c / (\rm{km}~\rm{s}^{-1})}{\nu /(\rm{kHz})}\\ 
 \Delta V_{R,n} &=& 8 \times 10^{-20}~W_{\nu}~T_{R,100}~n^{5.8}~\frac{c / (\rm{km}~\rm{s}^{-1})}{\nu /(\rm{kHz})}
\end{eqnarray}

Here $c$ is the speed of light, $\nu$ is the frequency, $T_{R,100}$ is the radiation temperature at 100~MHz and $W_{\nu}=\Omega/4\pi$ is the solid angle of the source on the sky. Whereas pressure broadening depends on $T_{e}$ and $n_{e}$, radiation broadening does not. Extrapolating the 408~MHz \citep{Ha82} and 150~MHz \citep{La70} radio continuum maps of the Milky Way, using a spectral index of 2.6, we find $T_{R(MW),100}\sim 2700$~K for the line of sight to Cyg~A. This corresponds to a FWHM velocity broadening of 9.6~km~s$^{-1}$ at 45~MHz ($n=535$). This is close to our measured value of 10 km~s$^{-1}$ and we therefore find that the Milky Way radiation field accounts for the bulk of the observed line broadening. 

In this case where radiation broadening dominates we can not obtain further useful constraints on $T_{e}$ and $n_{e}$ from pressure broadening. The range in $T_{e}$, $n_{e}$ values allowed for by the optical depth measurements also shows that pressure broadening is not significant and can at most contribute about 5~km~s$^{-1}$ to the FWHM velocity broadening of the line width at 45~MHz. For high signal-to-noise data it may be possible to distinguish the pressure contribution to the observed line width. However, with the current data set this is not possible. To explain the observed line widths with pressure broadening alone would require an electron density in excess of 0.1~cm$^{-3}$. This is inconsistent with the upper bound for $n_{e}$ obtained from the optical depths.

If all of the free electrons are from carbon then we can estimate the hydrogen density as $n_{H}=n_{e}/(1.5\times 10^{-4})\approx 400$~cm$^{-3}$ and the accompanying gas pressure as $p\sim n_{H}\times T_{e}\sim 4\times 10^{4}$~cm$^{-3}$~K. Here we used the gas phase abundance for carbon, i.e. [C/H]=1.5$\times$10$^{-4}$. The derived pressure is about an order of magnitude larger than the average pressure of the interstellar medium. Similar high pressures have previously been estimated for the CRRL emitting gas along the line of sight to Cas~A \citep[e.g. PAE89,][]{Ka98}. Enforcing pressure equilibrium between the CRRL gas and the average ISM, as advocated by \citet{Ka98}, would significantly reduce the allowed range in values for $T_{e}$ and $n_{e}$ for the CRRL emtting gas along the line of sight to Cyg~A. New observations, covering a wider range in frequency and higher sensitivity, will be needed to further constrain the CRRL emitting gas and investigate its pressure balance.

\subsection{An upper limit to the ionisation rate}\label{s_ionrate}
Having argued in Sect.~\ref{s_discuss_s1} that the observed RRL absorption is most likely due to carbon we can derive the total hydrogen ionisation rate limit $\zeta_{H}$ from our upper limit for hydrogen RRLs (HRRLs). This can be done in two ways. First by comparing our HRRL limit with H{\sc i} 21~cm emission. Following \citet{Sh76b} we can write,

\begin{eqnarray}
 \zeta_{H} &=& 5.7 \times 10^{-14}~\Phi_{2}~e^{\chi_{n}}~\frac{T_{e}^{2}}{T_{s}}~\frac{\nu /(\rm{GHz})}{(b_{n}\beta_{n})_{H}}~\frac{\tau_{H_{n}}}{\tau_{HI}}
\end{eqnarray}

We find $\zeta_{H}~<~$4$\times$10$^{-16}$~s$^{-1}$ for the range of $T_{e}$ values allowed by our optical depth measurements. Here we have used that $\tau_{\rm{peak,HI}}=0.36$ \citep{Ca98}, $n=535$, $\Phi_{2}$~=~3.34, $\nu$~=~45~MHz, ($b_{n}\beta_{n}$)$_{H}$~=~4.0 \citep{Po92}, $T_{s}=T_{e}$ and our 3$\sigma$ upper limit on $\tau_{\rm{peak,H_{n}}}$ (Table~\ref{t2_results}). $\Phi_{2}$ is an integral over gaunt factors \citep{Sp82}, $T_{s}$ is the H{\sc i} spin temperature and $(b_{n}\beta_{n})_{H}^{-1}$ are the departure coefficients for hydrogen.
 
We can also derive $\zeta_{H}$ by comparing our HRRL limit with our CRRL detection. Following \citet{Go09} we can write,

\begin{eqnarray}
 \zeta_{H} &=& 3.1 \times 10^{-15}~\Phi_{2}~\frac{n_{e}}{T_{e}^{0.5}}~\frac{(b_{n}\beta_{n})_{C}}{(b_{n}\beta_{n})_{H}}~\frac{\tau_{H_{n}}}{\tau_{C_{n}}}
\end{eqnarray}

Inserting our measured optical depths, ($b_{n}\beta_{n}$)$_{H}$~=~40.0 and ($b_{n}\beta_{n}$)$_{H}$~=~4.0 \citep{Po92} we again find $\zeta_{H}~<~$4$\times$10$^{-16}$~s$^{-1}$ for the allowed range of $T_{e}$ and $n_{e}$ values. Here we have used the gas phase abundance for carbon. If carbon is not depleted then the value of this upper limit goes up with a factor two to $\zeta_{H}~<~$8$\times$10$^{-16}$~s$^{-1}$.

The upper limit for the hydrogen ionisation rate derived here from RRLs is consistent with other measurements of the ionisation rate for the CNM by cosmic rays in our Galaxy \citep[e.g.][]{In07,Sh08}.

\subsection{Comparison with Cassiopeia A}
The most detailed low-frequency CRRL study to date is the one performed for the line of sight towards Cas~A \citep[e.g.][]{Ko81,Pa89,Ka98,St07,As13}. This line of sight, at $l$~=~111.74${\degree}$ and $b$~=~-2.14${\degree}$, shows CRRL emitting gas at a number of velocities that are consistent with individual clouds in the Orion (-5 to +5~km~s$^{-1}$) and Perseus (-50 to -35km~s$^{-1}$) spiral arms.

The velocity of the CRRL absorption observed by us is consistent with gas in the Orion spiral arm. The Orion arm CRRL component towards Cas~A is difficult to measure at low frequencies, because it is blended with the Perseus arm components. Therefore it has not been possible to model this component in detail. At 26 MHz, \citet{St07} estimates that the Orion arm component towards Cas~A has a peak optical depth $\tau_{peak}~\approx$~4$\times$10$^{-4}$, a line width FWHM~$\approx$~17~km~s$^{-1}$ and an integrated optical depth of about 0.6 in units of Hz. Upon extrapolating our best fitting model to 26~MHz we find a similar integrated optical depth for the CRRL emitting gas towards Cyg~A.

The Perseus arm CRRL component towards Cas~A is about an order of magnitude brighter than the Orion arm component at low frequencies. Therefore it has been measured with considerably better signal-to-noise and allows for a more detailed comparison. The best fitting model for this gas has T$_{e}~\approx$~75~K, n$_{e}~\approx$~0.02~cm$^{-3}$ and EM$_{C}~\approx$~0.01~~cm$^{-6}$~pc \citep{Ka98}. These parameters, in combination with the low value for the upper limit of the ionisation rate \citep{Pa89,As13} and the morphology, velocity structure of the CRRL emission, lead \citet{Ka98} to conclude that the CRRL emitting gas observed towards Cas~A is most likely associated with cold, diffuse, atomic clouds.

The measurements by \citet{Pa89}, \citet{Ka98} and \citet{As13} show that from 33 to 57 MHz the CRRLs for the Perseus arm component towards Cas~A decrease in linewidth by a factor 3 to 4 and decrease in integrated optical depth by about a factor 1.5 to 2, in units of Hz. This is similar to what we observe here and we therefore conclude that the CRRL absorption towards Cyg~A also likely originates in similar cold, diffuse, atomic clouds. 

\subsection{Cygnus A}
Hydrogen, helium or carbon RRL absorption associated with Cyg~A ($z=0.056$) itself was searched for in the redshift range 0.046 to 0.066 and constrained to be less than 1.5$\times$10$^{-4}$ at the 3$\sigma$ level for a channel width of 4~km~s$^{-1}$, see Table~\ref{t2_results}. This non-detection is not surprising as most of the low frequency radio flux from Cyg~A comes from its two giant radio lobes that extend far outside its host-galaxy (PGC~063932).

The optical size of the host-galaxy is between 15 and 30~arcsec. At the redshift of Cyg~A this corresponds to a physical size between 15 and 33~kpc. From a high resolution LOFAR map at 55~MHz (McKean et al. in prep.) we estimate that between 2 and 10 percent of the total radio continuum flux at this frequency comes from the optically emitting region. If we now assume that the interstellar medium of the host-galaxy is equal to its optical size then we find that our 3$\sigma$ RRL peak optical depth limit for the radio emission that likely traverses through the ISM of the host-galaxy is between 10$^{-3}$ and 10$^{-2}$.

The Milky Way is the only galaxy for which RRLs at frequencies below 1~GHz have been detected \citep[e.g.][EMA95]{Ka01}. At 76~MHz Erickson et al. find an average peak optical depth for carbon RRLs of a few times 10$^{-4}$ over large areas in the Milky Way plane. Cyg~A does not contain a significant amount of cold gas or starformation \citep*{Mc94}. Our non-detection of RRLs in this system is therefore not surprising.

\citet{Ka96} consider thermal absorption due to Galactic foreground gas, with a radio continuum optical depth $\tau_{74MHz}\sim0.3$ at 74~MHz, as an explanation for the spectral index assymetry measured for the two radio lobes of Cyg~A. The explanation they favor is that this is due to thermal absorption in foreground, extended HII region envelopes with n$_e~\sim$~5~cm$^{-3}$, T$_e~\sim$~8000~K and EM$~\sim$~2500~cm$^{-6}$~pc.

Our CRRL measurement traces gas that is cold and mostly neutral. Inserting our values, n$_e~\sim$~0.06~cm$^{-3}$, T$_e~\sim$~110~K and EM$~\sim$~0.001~cm$^{-6}$~pc, into formula 3.1 from \citet{Ka96} gives $\tau_{74MHz}~\sim$~3$\times$10$^{-5}$. We therefore find that the low-density, low-temperature CRRL gas observed here can not reproduce the required amount of thermal absorption.

\section[]{Conclusions}
In this paper we have presented high velocity resolution ($\sim$4~km~s$^{-1}$), low frequency ($\sim$45~MHz), stacked hydrogen, helium and carbon RRL spectra along the line of sight to Cyg~A taken with the LOFAR LBA array during its commissioning phase. 

\begin{itemize}
\item We have detected, for the first time, Milky Way foreground carbon RRLs in absorption against Cygnus~A.\\ 
\item The observed CRRL absorption has a peak optical depth of (2.2~$\pm$~0.5)$\times$10$^{-4}$, a FWHM line width of (10~$\pm$~3)~km~s$^{-1}$ and a local standard of rest velocity of (+4~$\pm$~1)~km~s$^{-1}$. The velocity of the gas is consistent with the Orion spiral arm in the Milky Way.\\
\item Our measurement allows for a large range in values for $T_{e}$ (50 to 500~K) and $n_{e}$ (0.005 to 0.07~cm$^{-3}$) of the associated gas. Our best estimate is $T_{e}~\sim$~110~K and $n_{e}~\sim$~0.06~cm$^{-3}$. These values are consistent with a CNM origin. Comparing to H{\sc i} 21~cm observations also shows a good match between H{\sc i} and CRRLs.\\
\item The RRL measurements provide an upper limit on the total ionisation rate of hydrogen $\zeta_{H}<$4$\times$10$^{-16}$~s$^{-1}$ for the CRRL emitting gas along the line of sight to Cyg~A.\\
\item RRLs from hydrogen, helium and carbon that are associated with Cygnus~A (z=0.056) itself, are constrained to have peak optical depths less than 1.5$\times$10$^{-4}$ at the 3$\sigma$ level for a 4~km~s$^{-1}$ channel.     
\end{itemize}

The optical depths of the CRRL absorption lines observed here along the line of sight to Cyg~A is similar to that seen along other sightlines (e.g. EMA95). However, our observations have an order of magnitude higher spatial resolution than these previous low-frequency CRRL absorption studies and show that we can use bright extragalactic radio sources for high resolution RRL pinhole studies of the interstellar medium. To the best of our knowledge this is the first time that low-frequency carbon RRLs associated with the Milky Way have been detected in absorption against an extragalactic radio source.

Future observations with LOFAR, using wider frequency coverage and higher sensitivity, will allow us to better constrain the properties of the CRRL emitting gas along the line of sight to Cyg~A and map its small-scale structure. In particular a deep ($\sim$8~h), high spectral resolution ($\Delta v\sim1~km~s^{-1}$) observation with the LOFAR High-band antenna (HBA) array in the frequency range 110-190~MHz will be helpful. Stacking LOFAR spectra from indivual sub-bands shows a decrease in spectral noise that is consistent with Gaussian noise behavior. This makes it a very promising technique for detecting faint RRLs with LOFAR.

\section*{Acknowledgments}
The authors would like to thank the LOFAR observatory staff for their assistance in obtaining and the handling of this large data set.

LOFAR, the Low Frequency Array designed and constructed by ASTRON, has facilities in several countries, that are owned by various parties (each with their own funding sources), and that are collectively operated by the International LOFAR Telescope (ILT) foundation under a joint scientific policy.

Support for this work was provided by NASA through Einstein Postdoctoral grant number PF2-130104 awarded by the Chandra X-ray Center, which is operated by the Smithsonian Astrophysical Observatory for NASA under contract NAS8-03060.

Chiara Ferrari and Giulia Macario acknowledge financial support by the {\it “Agence Nationale de la Recherche”} through grant ANR-09-JCJC-0001-01.

\bibliographystyle{mn2e}
\bibliography{rrl}

\vspace{0.3cm}
\noindent\hrulefill
\vspace{0.3cm}

\noindent $^{1}$ Netherlands Institute for Radio Astronomy (ASTRON), Postbus 2, 7990 AA Dwingeloo, The Netherlands\\ 
$^{2}$ Leiden Observatory, Leiden University, PO Box 9513, 2300 RA Leiden, The Netherlands\\
$^{3}$ Harvard-Smithsonian Center for Astrophysics, 60 Garden Street, Cambridge, MA 02138, USA\\
$^{4}$ Department of Physics and Astronomy, The Open University, Walton Hall, Milton Keynes MK7 6AA\\ 
$^{5}$ RAL Space, STFC Rutherford Appleton Laboratory, Chilton, Didcot, Oxfordshire OX11 0QX\\
$^{6}$ Space Telescope Science Institute, 3700 San Martin Drive, Baltimore, MD 21218, USA\\
$^{7}$ Astronomical Institute 'Anton Pannekoek', University of Amsterdam, Postbus 94249, 1090 GE Amsterdam, The Netherlands\\ 
$^{8}$ Max-Planck-Institut f\"{u}r Radioastronomie, Auf dem H\"ugel 69, 53121 Bonn, Germany\\
$^{9}$ SRON Netherlands Insitute for Space Research, PO Box 800, 9700 AV Groningen, The Netherlands\\
$^{10}$ Onsala Space Observatory, Dept. of Earth and Space Sciences, Chalmers University of Technology, SE-43992 Onsala, Sweden\\
$^{11}$ ARC Centre of Excellence for All-sky astrophysics (CAASTRO), Sydney Institute of Astronomy, University of Sydney Australia\\ 
$^{12}$ Centre for Radio Astronomy Techniques \& Technologies (RATT), Department of Physics and Elelctronics, Rhodes University, PO Box 94, Grahamstown 6140, South Africa\\
$^{13}$ SKA South Africa, 3rd Floor, The Park, Park Road, Pinelands, 7405, South Africa 
$^{14}$ Institute for Astronomy, University of Edinburgh, Royal Observatory of Edinburgh, Blackford Hill, Edinburgh EH9 3HJ, UK\\ 
$^{15}$ University of Hamburg, Gojenbergsweg 112, 21029 Hamburg, Germany\\
$^{16}$ Leibniz-Institut f\"{u}r Astrophysik Potsdam (AIP), An der Sternwarte 16, 14482 Potsdam, Germany\\ 
$^{17}$ School of Physics and Astronomy, University of Southampton, Southampton, SO17 1BJ, UK\\
$^{18}$ Research School of Astronomy and Astrophysics, Australian National University, Mt Stromlo Obs., via Cotter Road, Weston, A.C.T. 2611, Australia\\
$^{19}$ Max Planck Institute for Astrophysics, Karl Schwarzschild Str. 1, 85741 Garching, Germany\\
$^{20}$ Department of Astrophysics/IMAPP, Radboud University Nijmegen, P.O. Box 9010, 6500 GL Nijmegen, The Netherlands\\ 
$^{21}$ Th\"{u}ringer Landessternwarte, Sternwarte 5, D-07778 Tautenburg, Germany\\ 
$^{22}$ Hamburger Sternwarte, Gojenbergsweg 112, D-21029 Hamburg\\
$^{23}$ Laboratoire Lagrange, UMR7293, Universit\`{e} de Nice Sophia-Antipolis, CNRS, Observatoire de la C\'{o}te d'Azur, 06300 Nice, France\\ 
$^{24}$ Laboratoire de Physique et Chimie de l' Environnement et de l' Espace, LPC2E UMR 7328 CNRS, 45071 Orl\'{e}ans Cedex 02, France\\ 
$^{25}$ Jodrell Bank Center for Astrophysics, School of Physics and Astronomy, The University of Manchester, Manchester M13 9PL, UK\\ 
$^{26}$ Astronomisches Institut der Ruhr-Universit\"{a}t Bochum, Universitaetsstrasse 150, 44780 Bochum, Germany\\
$^{27}$ Astrophysics, University of Oxford, Denys Wilkinson Building, Keble Road, Oxford OX1 3RH\\
$^{28}$ Astro Space Center of the Lebedev Physical Institute, Profsoyuznaya str. 84/32, Moscow 117997, Russia\\ 
$^{29}$ Kapteyn Astronomical Institute, PO Box 800, 9700 AV Groningen, The Netherlands\\
$^{30}$ International Centre for Radio Astronomy Research - Curtin University, GPO Box U1987, Perth, WA 6845, Australia\\ 
$^{31}$ Center for Information Technology (CIT), University of Groningen, The Netherlands\\
$^{32}$ Centre de Recherche Astrophysique de Lyon, Observatoire de Lyon, 9 av Charles Andr\'{e}, 69561 Saint Genis Laval Cedex, France\\ 
$^{33}$ Fakult\"{a}t fur Physik, Universit\"{a}t Bielefeld, Postfach 100131, D-33501, Bielefeld, Germany\\
$^{34}$ LESIA, UMR CNRS 8109, Observatoire de Paris, 92195   Meudon, France\\
$^{35}$ Netherlands eScience Center, Science Park 140, 1098 XG, Amsterdam, The Netherlands\\
$^{36}$ Argelander-Institut f\"{u}r Astronomie, University of Bonn, Auf dem H\"{u}gel 71, 53121, Bonn, Germany

\newpage


\begin{table*}
 \centering
  \begin{tabular}{|l|l|} \hline
  Parameter & Value \\ \hline
  Data ID & L40787 \\
  Field center RA (J2000) & 19h59m28.3s \\
  Field center DEC (J2000) & +40d44m02s \\
  Observing date & 2011 December 27 \\
  Total on-source time & 10.5 h \\
  Frequency range & 33-57 MHz \\
  Number of sub-bands & 122~$^{a}$ \\
  Width of a sub-band & 0.1953 MHz \\
  Channels per subband & 512 \\
  Velocity resolution & 2.2-3.5 km~s$^{-1}$ \\ \hline
  \end{tabular}
 \caption[]{Details of the LOFAR LBA Observations. $^{a}$ 7 of these 122 sub-bands were removed from the analysis due to data corruption and an additional 10 sub-bands were removed due to strong RFI.}\label{t_obs}
\end{table*}

\begin{table*}
 \centering
  \begin{tabular}{|l|r|r|r|r|r|r|} \hline
  Stack & Range [MHz] & Range [n] & Centre [km~s$^{-1}$] & FWHM [km~s$^{-1}$] & $\int \tau$ dv [km~s$^{-1}$] & $\tau_{\rm{peak}}$ \\ \hline
  ALL & 33-57 & 487-583 & 3.8 $\pm$ 1 & 10.0 $\pm$ 3 & (2.1 $\pm$ 0.4) $\times$ 10$^{-3}$ & (2.2 $\pm$ 0.5) $\times$ 10$^{-4}$ \\
  LOW & 33-43 & 535-583 & 2.9 $\pm$ 1 & 19.0 $\pm$ 5 & (3.3 $\pm$ 0.8) $\times$ 10$^{-3}$ & (1.8 $\pm$ 0.6) $\times$ 10$^{-4}$ \\
  MID & 39-49 & 512-552 & 2.8 $\pm$ 1 &  9.8 $\pm$ 3 & (2.1 $\pm$ 0.6) $\times$ 10$^{-3}$ & (1.9 $\pm$ 0.6) $\times$ 10$^{-4}$ \\
  HIGH & 44-57 & 487-530 & 4.5 $\pm$ 1 & 5.8 $\pm$ 2 & (1.6 $\pm$ 0.5) $\times$ 10$^{-3}$ & (2.6 $\pm$ 0.7) $\times$ 10$^{-4}$ \\ \hline
  \end{tabular}
 \caption[]{Properties of the Milky Way foreground stacked carbon $\alpha$ line profile towards Cyg~A. The results for three different stacks are shown; (i) ALL: stack of all 48 subbands, (ii) LOW: sub-stack of the 21 lowest frequency carbon $\alpha$ lines, (iii) MID: sub-stack of 23 carbon $\alpha$ lines in the middle of the observed range, and (iv) HIGH: sub-stack of the 27 highest frequency carbon $\alpha$ lines. The line centre, FWHM and integrated $\tau$ values are obtained from a Gaussian fit to the spectrum. To convert $\tau$ from units of [km~s$^{-1}$] to [Hz] one may use that 1~km~s$^{-1}$ at 45~MHz corresponds to 150 Hz. The above value for $\tau$ thus corresponds to about 0.3 in units of [Hz] for the ALL stack. The peak optical depth is determined directly from the spectrum, see Fig.~\ref{f_results}.}\label{t_results}
\end{table*}

\begin{table*}
 \centering
  \begin{tabular}{|l|l|l|l|l|} \hline
  Source & Redshift (z) & $\tau_{\rm{peak}}$(HI $\alpha$) & $\tau_{\rm{peak}}$(HeI $\alpha$) & $\tau_{\rm{peak}}$(CI $\alpha$) \\ \hline
  Foreground & 0.0 & $<$1.4 $\times$ 10$^{-4}$ & $<$1.6 $\times$ 10$^{-4}$ & - \\
  Cygnus~A & 0.056 & $<$1.4 $\times$ 10$^{-4}$ & $<$1.2 $\times$ 10$^{-4}$ & $<$1.3 $\times$ 10$^{-4}$ \\ \hline
  \end{tabular}
 \caption[]{3$\sigma$ upper limits for the peak optical depth $\tau_{\rm{peak}}$ per 4~km~s$^{-1}$ channel of hydrogen, helium and carbon radio recombination lines along the line-of-sight to Cygnus~A. Values are given for both the foreground ($z=0$) and Cygnus~A ($z=0.056$).}\label{t2_results}
\end{table*}


\bsp

\label{lastpage}

\end{document}